\documentclass[]{interact}

\usepackage{enumitem}
\usepackage{footnote}
\usepackage{wrapfig}

\usepackage{epstopdf}
\usepackage[caption=false]{subfig}
\usepackage[natbibapa,nodoi]{apacite} 

\usepackage{float}
\usepackage{amsthm}
\usepackage{hyperref}
\hypersetup{%
  colorlinks=true,
  allcolors =blue
}
\usepackage{amssymb}
\usepackage{pifont}
\theoremstyle{plain}

\theoremstyle{definition}

\theoremstyle{remark}

\usepackage{tikz}
\usepackage{awesomebox}
\usepackage[nopar]{lipsum}
\usepackage{pifont}
\usepackage{amsmath}
\usepackage{dirtytalk}
\usepackage{xcolor}
\usepackage{multirow}

\definecolor{linkColor}{RGB}{6,125,233}
\definecolor{findingColor}{RGB}{172, 212, 227}

\usepackage[breakable, theorems, skins]{tcolorbox}
\usepackage{xcolor}
\definecolor{mygreen}{rgb}{0,0.5,0.2}

\definecolor{boxColor}{cmyk}{0.0, 0.1587, 1.0000, 0.0118, 1}
\definecolor{orange_own}{RGB}{255,159,0}

\makeatletter
\newcommand{\findingslabel}[2]{\def\@currentlabel{#2}\label{#1}}
\makeatother

\makeatletter
\newcommand{\keychallengelabel}[2]{\def\@currentlabel{#2}\label{#1}}
\makeatother

\newcounter{findingsCounter}

\newcommand{\revision}[1]{\textcolor{black}{#1}}

\begin{document}
\title{GazeBlend: Exploring Paired Gaze-Based Input Techniques for Navigation and Selection Tasks on Mobile Devices}
\author{
\name{Omar Namnakani\textsuperscript{a,*}\thanks{*Corresponding Author Omar Namnakani (omar.namnakani@gmail.com)}, Yasmeen Abdrabou\textsuperscript{b}, Jonathan Grizou\textsuperscript{a}\textsuperscript{,c}, and Mohamed Khamis\textsuperscript{a}}
\affil{\textsuperscript{a}University of Glasgow, Scotland, UK, \textsuperscript{b}Chair for Human-Centered Technologies for Learning, Technical University of Munich, Germany, \textsuperscript{c}GrizAI, UK}
}
\maketitle

\begin{abstract}
The potential of gaze for hands-free mobile interaction is increasingly evident. 
While each gaze input technique presents distinct advantages and limitations, a combination can amplify strengths and mitigate challenges. 
We report on the results of a user study (N=24), in which we compared the usability and performance of pairing three popular gaze input techniques: Dwell Time, Pursuits, and Gaze Gestures, for navigation and selection tasks while sitting and walking. 
Results show that pairing gestures for navigation with either Dwell time or Pursuits for selection improves task completion time and rate compared to using either individually.
We discuss the implications of pairing gaze input techniques, such as how Pursuits may negatively impact other techniques, likely due to the visual clutter it adds, how integrating gestures for navigation reduces the chances of unintentional selections, and the impact of motor activity on performance. Our findings provide insights for effective gaze-enabled interfaces.
\end{abstract}

\section{Introduction}
Gaze interaction on mobile devices is a growing research area exploring eye movements to control smartphones and tablets~\citep{khamis2018past}. It has the potential to become integral to daily device use~\citep{Møllenbach_Hansen_Lillholm_2013, 10.5555/1778331.1778385}. Advances in front-facing cameras and mobile processing power are revolutionising eye-tracking technologies~\citep{10.1145/3494999,huang2017tabletgaze, khamis2018past, 10.1145/3606947}, making gaze a promising input modality, especially when touch interaction is hindered due to contextual factors, such as when walking, carrying objects, in cold temperatures, or when the screen is wet~\citep{10.1145/3152771.3156161, 10.1145/1409240.1409253, goncalves2017tapping}.

\begin{figure}[t]
  \includegraphics[width=\textwidth]{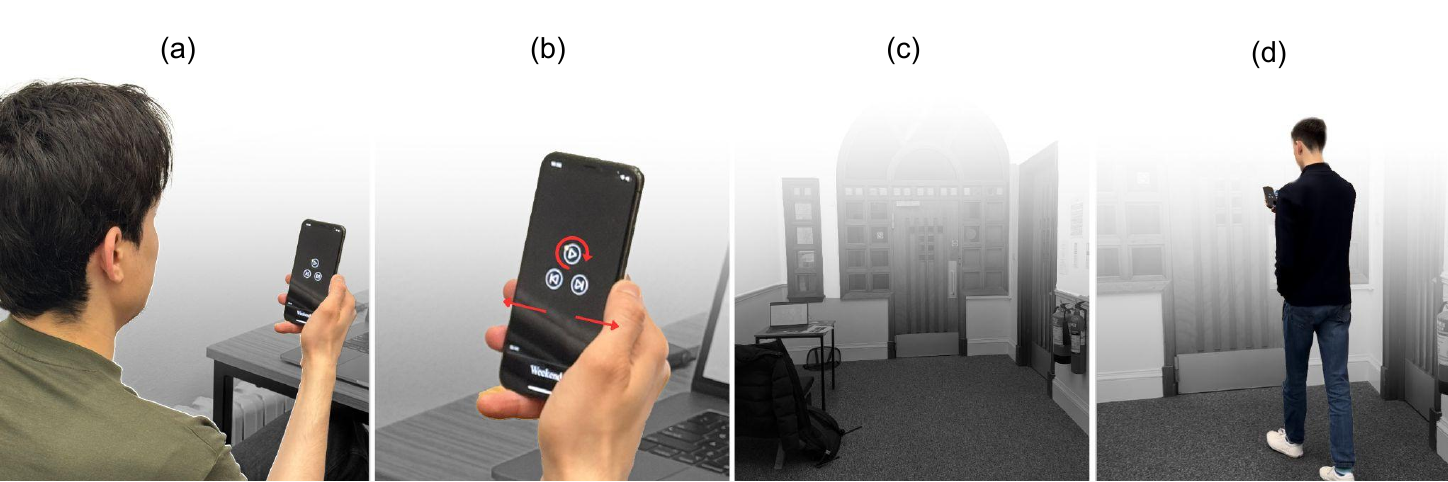}
  \caption{This work introduces and evaluates pairing gaze-based techniques: Dwell time, Pursuits, and Gestures for navigation and selection tasks on mobile devices, evaluating user perceptions within a single interface. We tested our techniques under two motor activities: while seated (a) and walking (d). (b) shows a participant navigating soundtracks via Gaze Gestures and selecting via Pursuits (arrows for illustration), and finally (c) shows the study location.}
  \label{fig:teaser}
\end{figure}

Prior research has examined gaze for hands-free mobile interaction~\citep{10.1145/3544548.3580871, kong2021eyemu, 10.1145/1378063.1378122, lei2023dynamicread}, exploring techniques based on distinct eye movements, such as Dwell time~\citep{10.1145/97243.97246, 10.1145/29933.275627}, Pursuits~\citep{VidalPusuits2013, 10.1145/2493432.2493477, 10.1145/2807442.2807499}, and Gestures~\citep{10.1145/1520340.1520699, 10.1145/2168556.2168602, Møllenbach_Hansen_Lillholm_2013}. The techniques have been evaluated in mobile contexts, such as walking~\citep{10.1145/3544548.3580871, 10.1145/1378063.1378122, lei2023dynamicread}. Each technique offers distinct advantages and limitations, with user preference and performance varying by context and task~\citep{10.1145/3544548.3580871, lei2023dynamicread}.
Dwell time, based on fixation, is well-known but demands accurate gaze estimation~\citep{khamis2018past, 10.1145/3419249.3420122, 10.1145/97243.97246,10.1145/3161168,10.1145/3287052} and limits the number of displayed targets to compensate for tracking inaccuracy~\citep{Møllenbach_Hansen_Lillholm_2013}.
Techniques that rely on relative eye movements, such as Pursuits and Gestures, are promising for mobile contexts~\citep{khamis2018past}. Pursuits can activate small targets~\citep{10.1145/2807442.2807499} but require dynamic interfaces~\citep{10.1145/3544548.3580871, VidalPusuits2013, 10.1145/2493432.2493477}. Gestures may be fast~\citep{10.1145/2168556.2168602} and work with fewer targets~\citep{10.1145/3544548.3580871}, but lack visual cues~\citep{Møllenbach_Hansen_Lillholm_2013}, rely on memorisation~\citep{10.1145/1743666.1743710}, and require more gestures as commands increase, complicating interaction~\citep{10.1145/1743666.1743710}.

Acknowledging each technique's advantages and limitations, we study how combining multiple gaze-based input techniques within the same interface can be leveraged to improve usability. We focus on pairing two gaze input techniques in the same interface, where we employ one technique for selection and another one for navigation. 
The topic is under-explored except for a few studies that split gaze-based selection into pointing and confirmation steps on a desktop computer, utilising different gaze input techniques for each~\citep{10.1145/2851581.2892291, 10.1145/3025453.3025455}.
Our work is the first to explore pairing gaze-input techniques for selection and navigation in general, and the first to examine the concept of pairing gaze input techniques on handheld mobile devices and in mobile settings, including while walking. 
We hypothesise that pairing gaze-based inputs, in this study, these are Dwell Time, Pursuits and Gestures, allows for effective, efficient and well-perceived interaction with mobile devices. 

To evaluate our concept, we conducted a user study ($N=24$) to assess the usability of six pairs of gaze-based techniques for selection and navigation tasks, in two motor activities: while sitting and while walking. 
We compared the pairs in terms of task completion time, task completion rates, error rates, and perceived workload using NASA-TLX. We also gathered participants' perceptions of pairing different eye movements through semi-structured interviews. 
In a task involving navigating a mobile home-screen-like interface to select a music player and play a given track, results show certain pairings improve the performance of gaze input, especially while walking. Using Gestures for navigation paired with Dwell time ($19.1s$) or Pursuits ($21.6s$) for selection was significantly faster than using Pursuits alone for both tasks ($27.4s$). Pairing also lowered error rates from ($57.1\%$) (Pursuits only) to $19.6\%$  (\textit{DwellGestures}) and $32.3\%$ (\textit{PursuitsGestures}), and significantly improved completion rates. 
Participants generally appreciated pairing techniques rather than using one eye movement during the interaction.

Our findings suggest that pairing gaze-based input techniques can improve the effectiveness of gaze interactions on mobile devices for navigation and selection tasks. 
Based on our findings, we discuss and present the implications of pairing gaze-based input techniques when designing user interfaces for enhanced gaze-based interaction on mobile devices. For example, we discuss the underlying mechanism that may impact the performance of such pairing.
We also discuss how, depending on the context, such as whether the user is walking or sitting, the studied techniques can be combined to maximise their strengths while minimising their limitations.

\section*{Contribution Statement}
The contribution of our work is threefold. Firstly, we introduce and investigate the novel concept of pairing gaze-based input techniques within the same interface. Secondly, we provide an in-depth analysis of this approach by (a) evaluating user performance and perception and (b) presenting implications for enhancing gaze interactions through pairing input techniques. Finally, we explore the effect of motor activity (sitting vs. walking) on user preferences and performance, offering valuable insights into the contextual adaptability of gaze-based interactions.

\section{Background and Related Work}
Gaze has long been studied for interaction. We build on insights from work that studied individual eye movements in isolation for explicit interaction or together with other eye movement types, with a focus on handheld mobile devices.

\subsection{Eye Movements and Gaze-based Interaction on Mobile Devices}
Previous research in the HCI domain has categorised the implementation of gaze-based interaction methods according to various eye movements, broadly, fixations, smooth pursuits, and saccades~\citep{Møllenbach_Hansen_Lillholm_2013, doi:10.1080/07370024.2017.1306444}.

Fixation refers to a state where the eyes appear motionless, though true fixations do not occur as microsaccades, drift, and tremors accompany them~\citep{Møllenbach_Hansen_Lillholm_2013, carpenter1988movements}. Dwell time~\citep{10.1145/97243.97246}, a fixation-based gaze input technique, is widely used for selecting static targets~\citep{10.1145/3544548.3580871, 10.1145/2168556.2168601, majaranta2019eye, li2021bayesgaze}. It involves a prolonged fixation to address the Midas touch problem~\citep{10.1145/97243.97246}, with durations ranging from under 200\,ms to over 1000\,ms~\citep{10.1145/3419249.3420122, 10.1145/1753846.1753983, majaranta2014eye, doi:10.1080/07370024.2017.1306444}. Dwell interfaces limit target numbers due to tracking inaccuracy and micro-movements~\citep{Møllenbach_Hansen_Lillholm_2013}. On mobile devices, dwell time has been compared with other gaze techniques~\citep{10.1145/3544548.3580871, lei2023dynamicread, 10.1145/2168556.2168601}, used for gaze-only interaction~\citep{10.1145/1378063.1378122}, and combined with other modalities~\citep{10.1145/1753846.1753983, kong2021eyemu}.

Smooth pursuits, a fixation-in-motion, occur when tracking a moving object~\citep{VidalPusuits2013, Møllenbach_Hansen_Lillholm_2013, 10.1145/3534624}. Pursuits input technique, based on this eye movements, compares the user's eye trajectory with that of the moving targets for selection~\citep{VidalPusuits2013, 10.1145/2807442.2807499, 10.1145/2971648.2971679}. With Pursuits, accuracy drops and demands increase with more targets~\citep{10.1145/2807442.2807499, 10.1145/3550301}. On mobile, Pursuits was the fastest compared with Dwell time and Gestures~\citep{10.1145/3544548.3580871} but was less preferred when walking due to higher demands~\citep{lei2023dynamicread}.

Saccades, rapid ballistic eye movements between fixations (30–120\,ms)~\citep{Møllenbach_Hansen_Lillholm_2013, doi:10.1080/07370024.2017.1306444, carpenter1988movements}, form the basis of gaze gestures~\citep{10.1145/2168556.2168602, 10.1145/3544548.3580871, 10.1145/2851581.2892291}. Gestures range from single strokes to multi-stroke patterns~\citep{10.1145/1520340.1520699, doi:10.1080/07370024.2017.1306444, majaranta2019eye}, can be fast, require no visual stimuli, and scale with more commands~\citep{10.1145/1743666.1743710, 10.1145/1520340.1520699, Møllenbach_Hansen_Lillholm_2013}. However, complexity reduces memorability, may cause fatigue, and simple gestures risk accidental activation~\citep{Møllenbach_Hansen_Lillholm_2013, 10.1145/3544548.3580871}. Gestures have been explored for mobile interaction~\citep{10.1145/2556288.2557040, 10.1145/2168556.2168601, rozado2015controlling, doi:10.1080/07370024.2017.1306444, lei2023dynamicread, 10.1145/3544548.3580871}. Since both Gestures and Pursuits techniques rely on relative eye movements, they do not require appropriate calibration as the absolute gaze positions are less important~\citep{10.1145/3025453.3025455, Møllenbach_Hansen_Lillholm_2013, khamis2018past}.

Overall, Prior work shows that mobile contexts present unique challenges~\citep{khamis2018past}. Thus, enabling gaze input on mobile devices has led recent research to explore gaze-input techniques in mobile settings~\citep{10.1145/3544548.3580871, lei2023dynamicread}, as findings from other settings, such as HMD~\citep{10.1145/3206505.3206522}, public displays~\citep{khamis2017uist}, and desktop~\citep{8311458} are not transferable to the context of mobile device.

\subsection{Combination of Gaze-based Input Techniques}
While there is work on combining gaze with other input modalities, such as mouse and/or keyboards~\citep{10.1145/332040.332444, 10.1145/1054972.1054994}, motion gestures~\citep{10.1145/3462244.3479938}, voice~\citep{10.1145/3490099.3511103} or touch input~\citep{10.1145/3317956.3318154, 10.1145/2984511.2984514, 10.1145/2807442.2807460, 10.1145/2642918.2647397, 10.1145/2207676.2208709}, in this work, we focus on pairing gaze-based input techniques in the same interface.
Prior work suggested combining gaze techniques to improve target selection, such as SPOCK and MSDT~\citep{10.1145/2851581.2892291, 10.1145/3025453.3025455}, which used Pursuits or Dwell time depending on whether stimuli were dynamic or static, with stimuli appearing after a 500\, ms dwell on a particular position. Ismoto et al.~used a two-level stroke gesture for command activation and a short dwell for target selection~\citep{isomoto2020gaze}, while Morimoto and Amir combined a short dwell to select a letter with a gesture across the opposite keyboard to activate it~\citep{morimoto2010context}. Since gestures can be performed without visual cues~\citep{Møllenbach_Hansen_Lillholm_2013}, it has been suggested, though not explored, that combining techniques such as Dwell time and gestures could strengthen gaze-based interaction~\citep{10.1145/2168556.2168579, 10.1145/1743666.1743710}.

As demonstrated, prior work mainly split a single interaction into steps using different techniques, rather than combining multiple gaze-based techniques to facilitate various interactions. While this addresses the Midas-touch problem, it can be limiting. We hypothesise that integrating multiple gaze techniques on mobile devices can improve performance and usability by enabling more natural interactions. 
To test this, we explore our research questions in mobile contexts, both seated and walking.

\subsection{Summary and Research Questions}
While prior work has used different eye movements for pointing and selection, no studies have examined pairing gaze-based techniques within the same interface for different functions. Existing combinations have focused only on improving target acquisition by dividing selection into pointing and confirmation.
Namnakani et al.~\citep{10.1145/3544548.3580871} compared various gaze-based input techniques when used independently for selection tasks within user interfaces and studied the impact of the number of targets on the techniques. We build on this research by investigating how pairing gaze input techniques for different functionalities on the same interface impacts the usability and user perceptions of the techniques. We study that in the context of mobile devices, where users interact with the device while sitting or walking.
Towards this end, we aim to answer the following questions: \textbf{RQ1:} How does pairing gaze-based techniques (Dwell time, Pursuits, Gestures) on a mobile interface affect performance, interaction efficiency and user perception compared to a single technique? And \textbf{RQ2:} How do motor activities (sitting vs. walking) influence performance and preferences with paired gaze-based techniques?
\section{Pairing Gaze-based Input Techniques Concept and Implementation}

To evaluate the usability of pairing different gaze-based techniques, we focus on two canonical user interface functions: \textit{selection}: commonly performed on mobile touch interfaces via tapping for discrete selection, and \textit{navigation}: typically executed using a single gesture by hand, i.e. swiping~\citep{10.1145/1743666.1743710} or tapping navigation buttons~\citep{Calleros2009Towards}. We employ one gaze technique for selection and another for navigation.
We developed an iPhone application using Xcode and Swift. Below, we list our implementation of the gaze-based inputs and how we pair them.

\subsubsection*{Dwell Time}
In our implementation, selection requires an 800\,ms fixation on the target. The dwell activation time was inspired by prior work~\citep{10.1145/3544548.3580871, 10.1145/3411764.3445177, 10.1145/3419249.3420122, 10.1145/1753326.1753645, niu2019improving}. 
The timer is fired when one gaze point crosses a dwell target, and a selection is performed when the user maintains their gaze within the boundaries of the target within the time window. Rather than treating each gaze point as individually landing inside the target, we calculate the mean fixation point within each time window and determine if it lies inside the target, to compensate for tracking inaccuracy in the mobile context~\citep{10.1145/3544548.3580871}.

\subsubsection*{Pursuits}
Our Pursuits implementation follows Namnakani et al. and Esteves et al.~\citep{10.1145/2807442.2807499, 10.1145/3544548.3580871}. As Pursuits requires dynamic interfaces~\citep{10.1145/2493432.2493477, 10.1145/2807442.2807499}, users select targets by following moving objects. On the mobile’s limited display, targets are shown as static circles, each with a small orbiting circle. Selection occurs by following the orbiting circle of the desired target~\citep{10.1145/2807442.2807499, 10.1145/3544548.3580871}. We use the Pearson correlation coefficient~\citep{10.1145/3544548.3580871, 10.1145/2807442.2807499, 10.1145/2493432.2493477, 10.1145/3206505.3206522, 10.1145/3064937} to compare each target’s horizontal and vertical trajectory with the x and y coordinates of the gaze data within a time window of one second (30-sample)~\citep{10.1145/3544548.3580871, esteves2020comparing, 10.1145/2807442.2807499}. The lower correlation value between x and y is stored, and a target is selected if the stored correlation exceeds a threshold of $0.8$, following practices used in prior work \citep{10.1145/3544548.3580871, 10.1145/2807442.2807499}. If not, correlation is recalculated with each new sample in a rolling window of 30 samples~\citep{10.1145/3544548.3580871, 10.1145/2807442.2807499}. All orbiting circles move at 120°/sec, with half moving in opposite directions and all starting at different positions, following~\citep{10.1145/3544548.3580871, 10.1145/2807442.2807499}.

\subsubsection*{Gaze Gestures}
We used single-stroke right and left gestures, as they are efficient for basic navigation~\citep{Møllenbach_Hansen_Lillholm_2013, 10.1145/1520340.1520699}. 
A gesture had to be completed within 1000\,ms (30 samples), with the Pearson correlation coefficient~\citep{10.1145/3544548.3580871} used to compare the user’s horizontal gaze path with gesture templates. 
\revision{Consistent with prior work~\citep{10.1145/3544548.3580871}}, the template with the highest correlation $> 0.8$ was taken as the intended gesture. 
\revision{To prevent unintentional gaze gesture activation by distinguishing natural horizontal eye movements from navigation gestures, we employed the screen bounds gazing to complete them, similar to prior work \citep{10.1145/3544548.3580871, 10.1145/2168556.2168579}. 
By incorporating a 160 pixels-wide horizontal buffer at both the left and right screen edges, spanning the entire length of the device~\citep{10.1145/3544548.3580871}, the gesture is activated once the final horizontal gaze point crosses the buffer area, signaling completion of a gesture~\citep{10.1145/3544548.3580871, 10.1145/2168556.2168579, 10.1145/1520340.1520699}.}

\begin{figure}[t]
    \centering
    \includegraphics[width=0.55\textwidth]{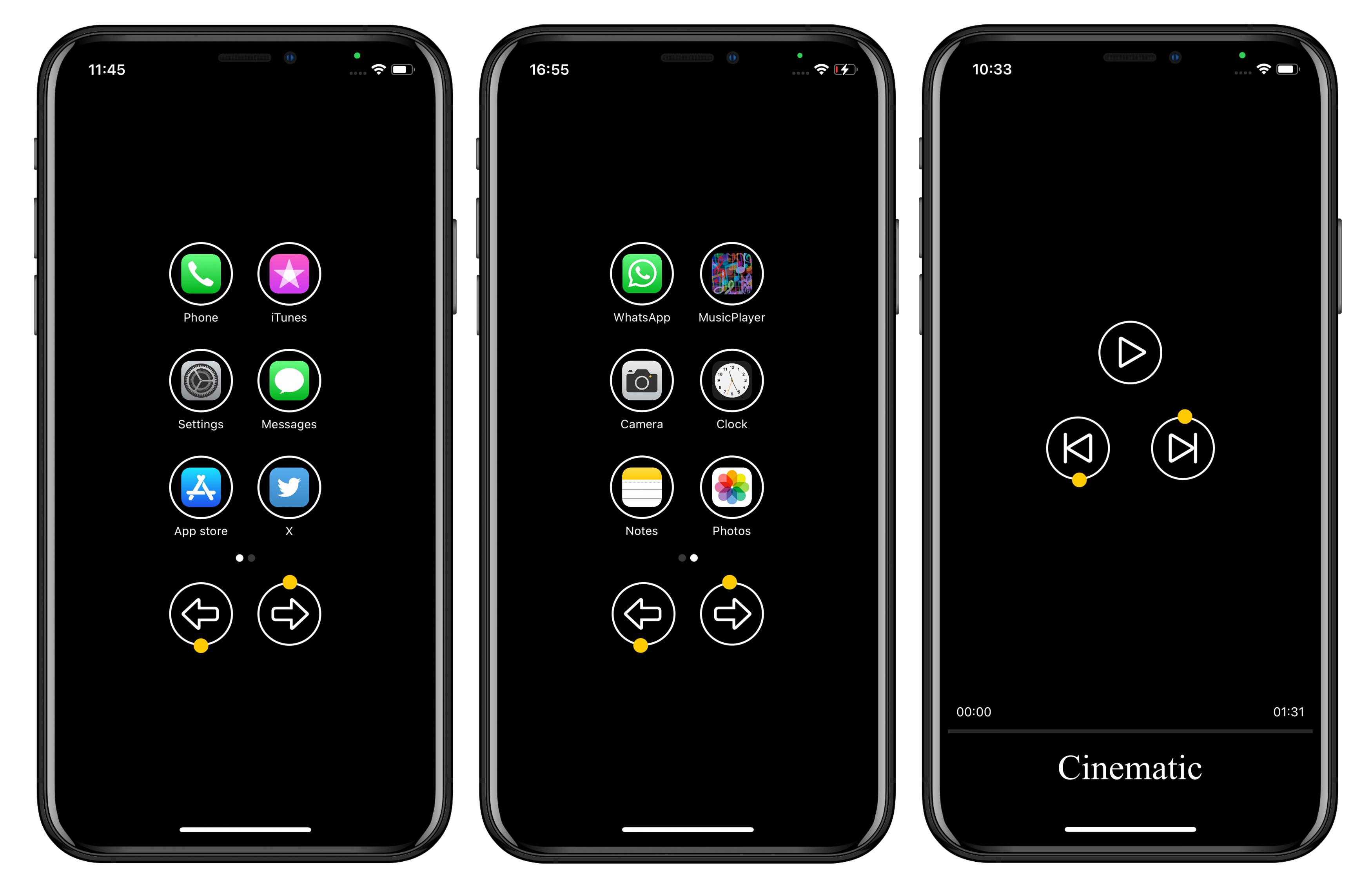}
    \caption {A step-by-step showing how participants perform the task for the \textit{DwellPursuits} condition. In the \textit{DwellPursuits} condition, Dwell time handled selection and Pursuits handled navigation: (1) navigate from page 1 (left) to 2 (middle) via Pursuits in the home screen, (2) select the music player with Dwell time (middle), (3) navigate four screens to the target soundtrack via Pursuits (right), and (4) play it with Dwell time (a selection step). Navigation buttons remain on-screen for consistency, though unused with Gestures.
    }
    \label{fig:screenshot}
    \vspace{-3mm}
\end{figure}

\subsubsection*{Pairing Gaze-based Techniques Prototype} \label{prototype}
We implemented a prototype application with different pairings of gaze-based techniques (see Figure \ref{fig:screenshot}). The interfaces mimic typical Android and iOS layouts. The home screen shows six selectable apps and two navigation targets (left and right). The music player provides three controls: play/pause, next, and previous. Although gestures can be used without visual targets~\citep{Møllenbach_Hansen_Lillholm_2013}, all targets, including navigation buttons, were made visible for consistency. The number of home screen targets was limited to eight, as prior work found higher error rates with Dwell time and Pursuits when more targets are displayed~\citep{10.1145/3544548.3580871, 10.1145/2807442.2807499}.
We used a circular target with a diameter of 65\,pt (1.08 cm or 2.09° at 29.6 cm viewing distance, measured with Eyedid during the experiment), corresponding to 195 pixels on the iPhone X used in the study.
Our decision follows Apple design guidelines~\citep{apple_guide} and is inspired by prior work on gaze interaction on mobile devices~\citep{10.1145/3544548.3580871}. Targets were positioned closer to the centre of the display as previous work reported reduced accuracy toward screen edges~\citep{10.1145/2785830.2785869, park2021gazel, 10.1145/3544548.3580871}.
When integrating Pursuits and Gestures, we used iOS Swift’s concurrency features, such as DispatchQueues and DispatchGroup, to process gaze data for each technique concurrently within a 1000\,ms window. 
DispatchQueue handles the parallel execution, calculating correlations and returning the target with the highest correlation that exceeds a $0.8$ threshold. DispatchGroup ensures that all processes are complete before selecting and executing the target. If both techniques exceed the threshold, we choose the target with the highest correlation. 
To integrate Dwell Time with Pursuits or Gestures, we run each technique's processes concurrently on separate queues, while sharing a state that is updated upon input detection. Using DispatchQueue, we ensured exclusive access to this state. Upon detecting input, all processes are reset, and data collection starts again.
\section{Evaluation}
We conducted our study in a low-traffic corridor of a university building to examine the usability of pairing gaze-based selection and navigation tasks. The study was approved by the University of Glasgow, College of Science and Engineering Ethics Committee (Approval No. 300220210).

\subsection{Study Design}
To study the impact of gaze-based input pairing in the same interface, we designed a $6x2$ within-subjects factorial design with the following two independent variables. \textbf{[IV1] Input techniques:} We tested six technique pairs: 1) Dwell time and Dwell time, 2) Dwell time and Pursuits, 3) Dwell time and Gestures, 4) Pursuits and Pursuits, 5) Pursuits and Dwell time, and 6) Pursuits and Gestures (see table~\ref{tab:combinations}). In each pair, the first technique was used for selection (e.g., activating the music player or playing a soundtrack), and the second for navigation (e.g., browsing apps or changing soundtracks). \textbf{[IV2] Motor activity:} We covered sitting and walking conditions. 

\begin{table}[t]
\caption{The six conditions of gaze input techniques used for selection and navigation.}

\label{tab:combinations}
\begin{tabular}{@{}lllll@{}}
\toprule
\multicolumn{2}{l}{\multirow{2}{*}{}}    & \multicolumn{3}{c}{Navigation}    \\
\multicolumn{2}{l}{}                     & Dwell Time & Pursuits & Gestures \\ \midrule
\multirow{3}{*}{Selection} & Dwell Time & \textit{DwellDwell}          & \textit{DwellPursuits}        & \textit{DwellGestures}        \\
                            & Pursuits   & \textit{PursuitsDwell}          & \textit{PursuitsPursuits}        & \textit{PursuitsGestures}        \\
                            & Gestures   & -          & -        & -        \\ \bottomrule
\end{tabular}
\vspace{-4mm}
\end{table}

\subsection{Recruitment and Participants}
Through social media, word of mouth, and posters across campus, we recruited $24$ participants for the study (13 male, 11 female; ages 18–52, $M=28.87$, $SD=8.0$, $22$ right-handed).
Although none of the participants wore eyeglasses during the experiment, six reported nearsightedness, while three indicated they are farsighted.
The participants indicated moderate use of phones while walking ($M=3.08, SD=1.10$) on a 5-point scale (0: never to 5: always). They also reported little to no experience with eye-tracking ($M=0.96, SD=1.63$) on a 5-point scale (0: no experience to 5: very experienced). The participants received a compensation of £10.

\subsection{Apparatus and Task} \label{task}
We used an iPhone X running iOS 16.6 to run the prototype described in Section \ref{prototype}. The device features a front-facing camera with a 7-megapixel sensor. Eyedid SDK~\citep{seeso} was used to enable eye tracking on the mobile device. \revision{With a gaze tracking rate of 30 Hz (30 frames per second), the SDK provides real-time gaze estimation data in the form of (x,y) screen coordinates and reports an accuracy of 1.6°. The (x,y) values are provided even when the estimated gaze position falls outside the screen boundaries.}

The Music player prototype consisted of 30 soundtracks downloaded from \href{http://www.pixabay.com}{Pixabay.com} and ordered randomly in each task. We chose random soundtracks without showing the artists' names, rather than songs, to minimise the participants' familiarity with the songs. To avoid the possibility that participants might forget the soundtracks' titles during the tasks, we referred to them only by single-word titles. 

The main task is to play a specific soundtrack given to the participant. This is conducted through four main steps:

\begin{enumerate}[
  label=Step~\arabic*:,   
  align=left,             
  labelsep=.5em,          
  labelwidth=*,           
  labelindent=10pt,        
  leftmargin=*            
]
 \item (Navigation), Swipe through a Home screen-like application to find the music player prototype (see Figure \ref{fig:screenshot}).
 \item (Selection), Open the music player prototype.
 \item (Navigation), Swipe through soundtracks to find the given soundtrack.
 \item (Selection), Play the given soundtrack once found.
\end{enumerate}

In the sitting condition, participants sat in a relaxed posture. In the walking condition, participants were instructed to walk as they normally would when using a mobile device, holding it in whichever hand felt natural. During walking, they were guided to walk the length of the corridor from one end to the other, travelling back and forth without stopping. The experimenter monitored participants to ensure continuous movement during task performance.

\subsection{Measures}
To evaluate our input technique pairings, we measured task completion time, completion rate, error rate, and perceived workload (NASA-TLX~\citep{doi:10.1177/154193120605000909}). Completion time is calculated as the duration from the Home screen to playing the correct soundtrack. Completion rate is calculated as the percentage of tasks completed within 60 seconds. Error rate is calculated as the proportion of incorrect actions to total actions (e.g., 8 errors in 15 actions = 53.3\%). We also collected qualitative feedback through semi-structured interviews and Likert-scale questions on usability.

\subsection{Procedure} \label{sec:procedure}
Upon arrival, participants were briefed, signed the consent, and provided demographics, including eye-tracking experience. They received a demonstration of the interface and task, then completed the study tasks (Section~\ref{task}). A calibration phase preceded all techniques to ensure consistent tracking, even for Pursuits and Gestures where it was not required~\citep{10.1145/3544548.3580871, khamis2018past, 10.1145/2168556.2168601, 10.1145/2493432.2493477, 10.1145/3229434.3229452}.
Prior to their participation in the study, participants were shown the designated location and the specific path they were to follow during the walking tasks to acquaint them with the environment.
\textbf{Before each condition,} participants viewed a screen containing detailed information about the task to be performed, including the gaze input techniques to use, the name of the application to open, the soundtrack's title to play, and whether to perform the task sitting or walking. 
\textbf{In each condition}, \revision{participants completed one training trial, which was similar to the main task, with the only difference being the soundtrack titles, to familiarise themselves with the task and techniques.} 
To focus on selection performance rather than search time~\citep{10.1145/3544548.3580871}, the music player was placed in the same location for all runs (steps 1 and 2). Training helped participants locate the app so search time would not affect analysis. Soundtrack titles for steps 3 and 4 were randomly chosen but consistently positioned. Participants completed three trials per condition, each requiring two selections (activating the music player and playing the soundtrack) and five navigation steps (one to move to the second home-screen page and four to find the soundtrack), see Figure~\ref{fig:screenshot}.
Incorrect selections of apps triggered a one-second alert and were logged, while playing incorrect soundtracks were logged silently. 
\revision{The trial began when the participant was presented with the Home screen in the prototype and ended when the correct soundtrack was played, prompting a completion message.}
\revision{Trials were unsuccessful and also ended if the correct soundtrack was not played within 60 seconds or if the next correct action was not taken within 20 seconds at any stage~\citep{10.1145/3544548.3580871}. This approach was adopted to limit the experiment duration and minimise participant fatigue. The starting and ending times of each trial were automatically logged using timestamps from the phone.}
Feedback is provided to the participant whenever a button is activated by changing the button's colour to yellow for one second. 
\textbf{After each condition,}  participants completed the NASA-TLX questionnaire and a 5-point Likert scale to reflect on various usability aspects. \textbf{At the end of the experiments,} participants filled out a questionnaire to rank the paired techniques based on their preference, and we conducted a semi-structured interview. Participants completed a total of 36 trials. All conditions were counterbalanced using Latin square to counteract order effects.

\subsection{Limitations} \label{sec:limitation}
\revision{In our study, participants located the music player app during the training step (see Section \ref{sec:procedure}), thereby eliminating the need for a search step during the main tasks. 
While this design decision may not reflect real-world scenarios, it was essential to minimise search time, which could vary regardless of the input technique, as we did not want longer searches to be mistaken for slower gaze input. 
We did not study the impact of the search phase on gaze techniques as this was already explored in prior work by Namnakani et al.~\citep{10.1145/3544548.3580871}. 
The researchers explored Dwell time, Pursuits, and gestures for target selection using various numbers of targets in two motor activities: sitting and walking. 
In a task in which participants had to locate and select discrete targets, gaze gestures was reported as the most accurate input technique. Dwell time was more accurate than Pursuits as the number of targets increased, especially when walking.}

Although participants in our experiment did not face obstacles typical of real-world settings, conducting the study in a corridor may have introduced partially controllable, yet unpredictable, factors. For example, distractions such as people passing through or interacting with the experimenter could momentarily disrupt concentration; however, no such incidents occurred, and we were prepared to repeat any trials that were notably disrupted if necessary. We intentionally chose an indoor environment with minimal distractions to evaluate the performance of pairing the techniques in isolation, yet this may not fully reflect the wide range of real-world contexts where gaze interaction can be used, such as outdoor environments, crowded spaces, or varying lighting conditions. While these contextual factors can be investigated, in our experiment, we controlled them to mitigate the effects of confounding variables on results, while still ensuring that the study was conducted under natural usage conditions to maximise the relevance and applicability of our findings.

\revision{We conducted the study on an iPhone X, released in 2017. Since pairing gaze techniques required running multiple threads concurrently to detect inputs via Dispatch Queues, this may have introduced additional computational load due to the hardware's age. While we did not explicitly monitor CPU usage or latency, we did not observe any latency or delays during the pilot testing and the experiment, nor was it reported as such by participants.
While performance may vary across newer mobile devices, in our study, all paired techniques were evaluated under identical hardware and software conditions in the same controlled environment. Thus, any performance limitations are expected to have affected all techniques equally.}

Finally, our results are based on tracking data from the Eyedid library on mobile devices. While accuracy may vary with other software, our focus was on the conceptual understanding of user interfaces, specifically, how participants perceive paired gaze input techniques. Prior work suggests that tracking quality is influenced more by individual user differences than by the tracker or lighting~\citep{10.1145/3025453.3025599}.

\section{Results}
For the statistical tests, unless specified otherwise, we analysed the collected measures using repeated measures ANOVA with post-hoc pairwise comparisons and Bonferroni correction. For sphericity violations, we report Greenhouse-Geisser-corrected degrees of freedom and p-values. Where data transformation was required, we applied the Aligned Rank Transform (ART)~\citep{10.1145/1978942.1978963}. If participants failed to complete all the trials within 60 seconds for a condition, task time was replaced with the mean of the remaining values for that condition~\citep{10.1145/3544548.3580871}. Due to a device error, one trial for one participant in one condition was lost, leaving two trials for that condition and a total of $24\times12\times3-1=863$ measurements.

\begin{figure}[!t]
    \centering
    \includegraphics[width=\linewidth]{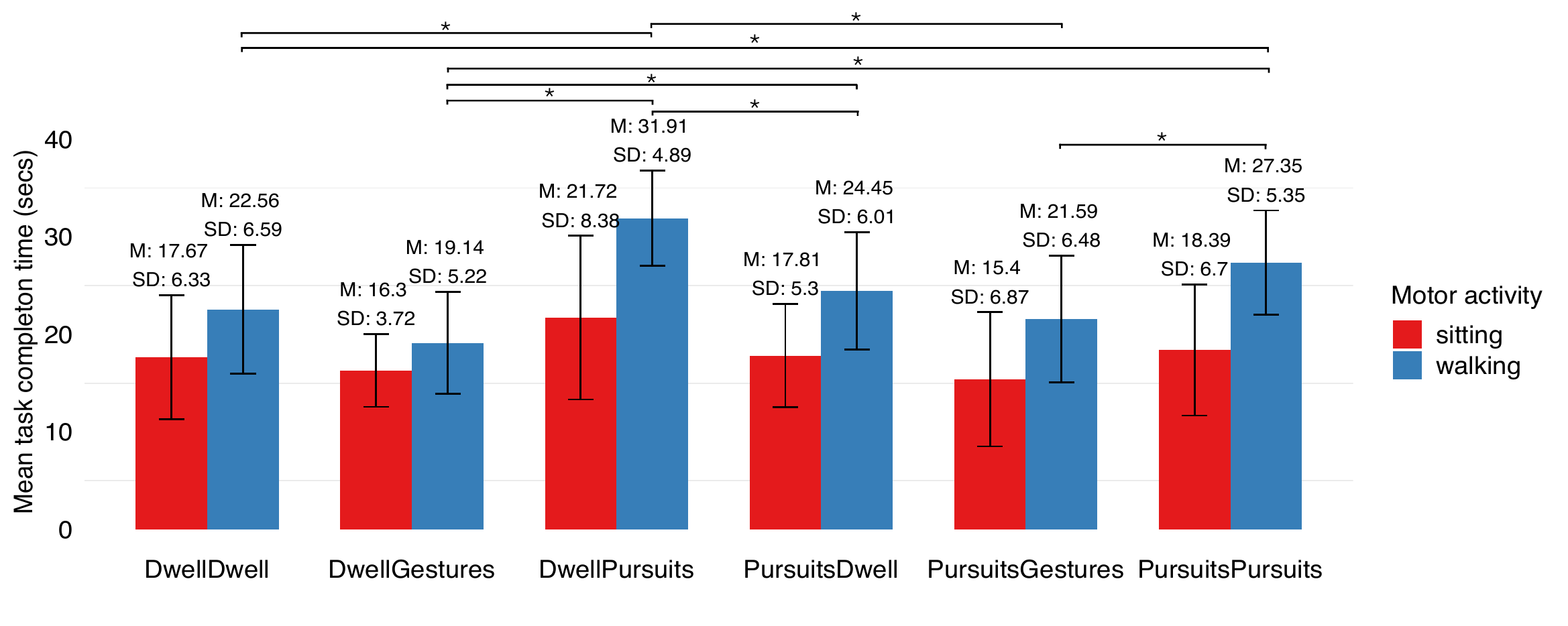}
    \caption{This figure represents the mean task completion time in seconds. No significant differences were observed between the techniques during seated motor activity. While walking revealed significant differences between several pairs, \textit{DwellGestures} and \textit{PursuitsGestures} showed faster completion time than \textit{PursuitsPursuits}.
    The error bars represent the standard deviation. }
    \label{fig:completion_time}
    \vspace{-3mm}
\end{figure}

\subsection{Task Completion Time}
Figure \ref{fig:completion_time} summarises the descriptive statistics and shows the significant differences between pairs. 
The statistical test revealed a two-way interaction between Input techniques and Motor activity on task completion time, F\textsubscript{5,115}=2.79, $p<0.05$. While we found no significant differences between the input techniques in the sitting motor activity on the task completion time, when walking, the post hoc analysis showed that \textit{DwellGestures} ($M= 19.1s, SD= 5.22$, $p<.001$) and \textit{PursuitsGestures} ($M= 21.6s, SD= 6.48$, $p<.05$) are significantly faster than \textit{PursuitsPursuits} ($M= 27.4s, SD= 5.35$). 
\textit{DwellGestures} also had significantly faster completion time than \textit{DwellPursuits} ($M= 31.91s, SD= 4.9$, $p<.0001$) and \textit{PursuitsDwell} ($M= 24.5s, SD= 6.01$, $p<.05$). 
Using t-test, we also looked at the impact of the motor activity on the techniques' completion time. We found that \textit{DwellDwell}, \textit{DwellPursuits}, \textit{PursuitsDwell}, \textit{PursuitsGestures}, and \textit{PursuitsPursuits} are significantly slower when performing the task while walking (all $t(23) <= -2.867, p < .01$). This implies that there is not enough evidence to suggest that changing the motor activity results in additional time when using \textit{DwellGestures}, $t(23) <= -2.008, p > .05$. 

Overall, \textit{DwellGestures} reduces task completion time without clear evidence of whether the motor activity impacts its performance.

\subsection{Task Completion Rate}
\begin{figure}[!t]
    \centering
    \includegraphics[width=\linewidth]{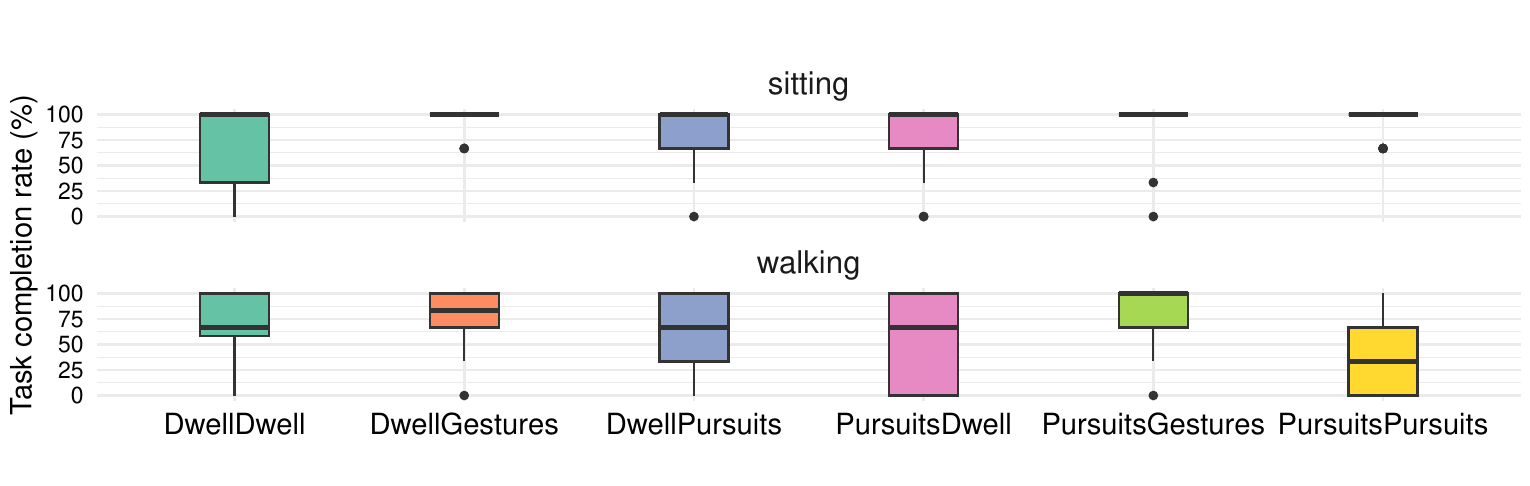}
    \caption{The figure illustrates the task completion rate by input techniques and motor activities. 
    \textit{DwellGestures} and \textit{PursuitsGestures} resulted in a significantly higher completion rate compared to \textit{PursuitsPursuits} when walking.}
    \label{fig:completion_rate}
    \vspace{-3mm}
\end{figure}

Figure \ref{fig:completion_rate} summarises the descriptive statistics. The statistical test revealed a two-way interaction between input techniques and motor activity on task completion rate, F\textsubscript{5,253}= 5.13, $p< .001$. 
While the Friedman test revealed a statistically significant variation in the task completion rate between the input techniques while sitting, $\chi\textsuperscript{2}(2)$ = 18.05, $p=.003$, post hoc analyses using Wilcoxon signed-rank tests, adjusted with a Bonferroni correction, did not show notable differences between the pairs.
When walking, there were significant differences between the techniques, $\chi\textsuperscript{2}(2)$ = 18.093, $p=.003$. Post hoc analysis revealed significant differences between \textit{PursuitsPursuits} ($Med= 33.3\%, IQR= 66.7\%$) and both \textit{PursuitsGestures} ($Med= 100\%, IQR= 33.3\% $) ($p<.01$) and \textit{DwellGestures} ($Med= 83.3\%, IQR= 33.3\%$) ($p<.05$). 
This suggests that pairing gestures with either Dwell time or Pursuits significantly enhances the task completion rate compared to using Pursuits alone.

\subsection{Task Error Rate}
\begin{figure}[!t]
    \centering
    \includegraphics[width=\linewidth]{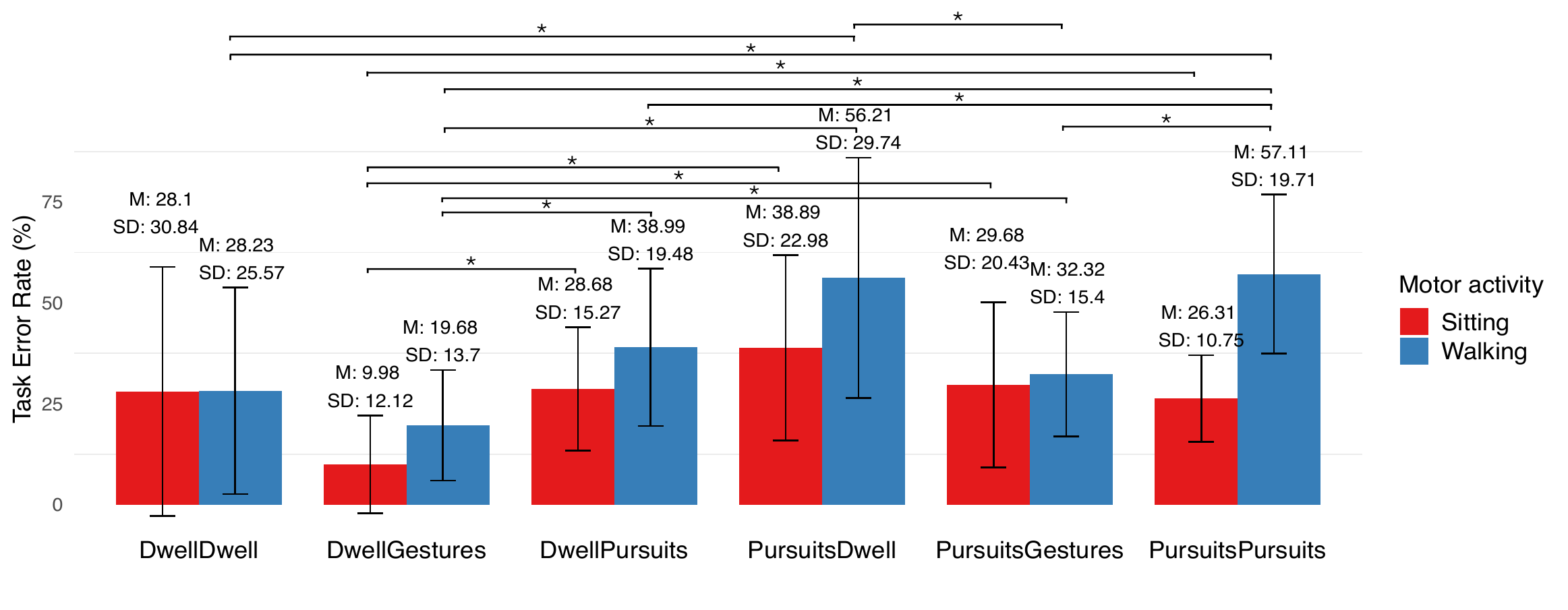}
    \caption{Sitting and walking task error rates. For tasks performed while seated, \textit{DwellGestures} demonstrated significantly higher accuracy than \textit{DwellPursuits}, \textit{PursuitsDwell}, \textit{PursuitsGestures}, and \textit{PursuitsPursuits}. In contrast, during walking, \textit{DwellDwell}, \textit{DwellGestures}, and \textit{PursuitsGestures} significantly outperformed \textit{PursuitsPursuits} in terms of accuracy. The error bars represent the standard deviation.}
    \label{fig:error_rate}
\end{figure}

Figure \ref{fig:error_rate} summarises the descriptive statistics and details significant differences between multiple pairs. 
We found a statistically significant two-way interaction between Input techniques and Motor activity on task error rate, F\textsubscript{3.37, 77.52}= 3.83, $p<0.05$.
When seated, post hoc analysis revealed that \textit{DwellGestures} is significantly less error-prone ($M= 9.98\%, SD= 12.1$) compared to \textit{DwellPursuits} ($p<.01$), \textit{PursuitsDwell} ($p<.01$), \textit{PursuitsGestures} ($p<.05$), and \textit{PursuitsPursuits} ($p<.0005$). 
When walking, \textit{PursuitsGestures} is significantly more accurate ($M= 32.3\%, SD=15.4$) compared to \textit{PursuitsPursuits} ($M= 57.1\%, SD= 19.7$). 
The analysis shows that \textit{DwellGestures}, \textit{DwellPursuits}, \textit{PursuitsDwell} (all $p <.05$) and \textit{PursuitsPursuits} ($p< .0001$) are substantially affected by motor activities. 

Overall, \textit{DwellGestures} led to a significantly lower error rate than any paired techniques involving Pursuits when sitting. When walking, the pairing of gestures with either Dwell time or Pursuits proved to be significantly more accurate than \textit{PursuitsPursuits}.

\subsection{Task-load Index}
\begin{figure}[!t]
    \centering
    \includegraphics[width=\linewidth]{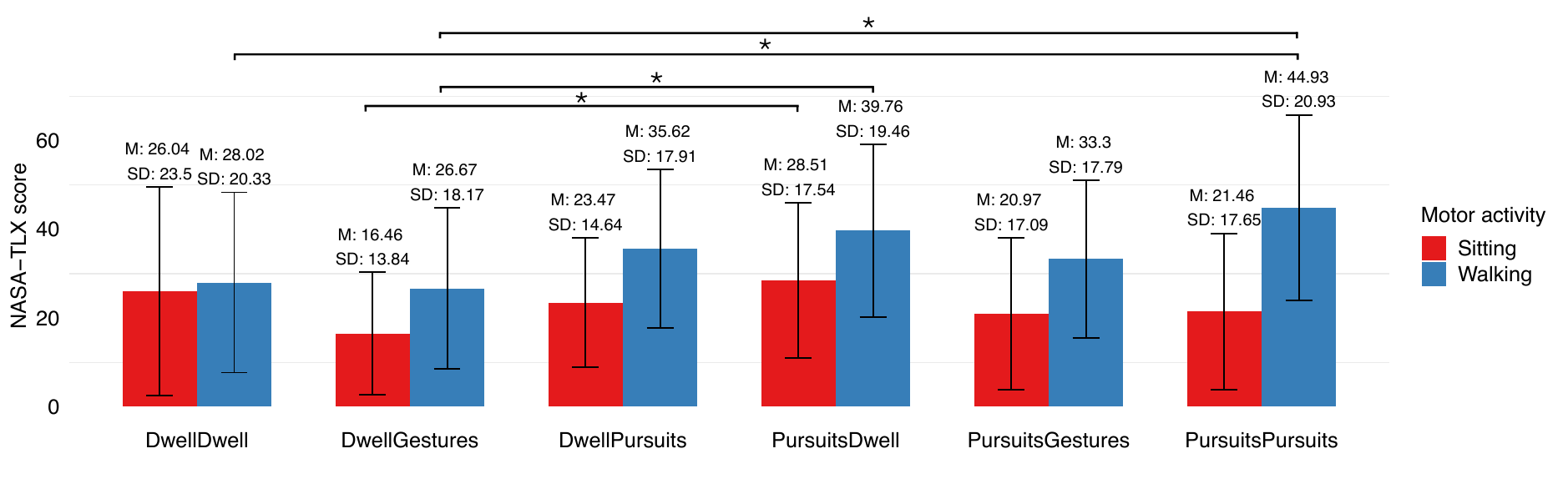}
    \caption{Mean overall NASA-TLX workload per input techniques in each motor activity. The error bars represent the standard deviation.}
    \label{fig:nasa_overall}
\end{figure}

We calculated mean TLX scores across all six dimensions for the twelve conditions using the raw NASA-TLX method~\citep{doi:10.1177/154193120605000909}. The scores are based on a scale of 100~\citep{10.1145/3544548.3580871, 10.1145/3411764.3445478, 10.1145/3290605.3300521, 10.1145/3173574.3174221}. The lower the scores, the lower the workload (See Figure \ref{fig:nasa_overall}).
There was a statistically significant interaction between Input techniques and motor activity on the task load, F\textsubscript{5,253}= 2.87, $p< .05$. 
Post hoc analysis showed that \textit{DwellGestures} resulted in significantly lower workload than \textit{PursuitsDwell} while sitting ($M=16.5, SD=13.8$ compared to $M=28.5, SD=17.5$, $p<.05$) and walking ($M=26.7, SD=18.2$ compared to $M=39.8, SD=19.5$, $p<.05$).
\textit{PursuitsPursuits} showed a significantly higher cognitive workload ($M= 44.9, SD= 20.9$) compared to both \textit{DwellDwell} ($M= 28.0, SD= 20.3$) and \textit{DwellGestures} when walking ($p<.001$).

\subsection{Likert Scales \& Preference Rankings} \label{likert_section}
Participants used a 5-point Likert scale (1= Strongly Disagree; 5= Strongly Agree), to answer questions that evaluated various aspects of navigation and selection using different eye movements. 
We structured questions as statements such as "I found navigating with \{ Input Technique \} and selecting with \{ Input Technique \} to be easy, accurate, easy to learn, enjoyable, not requiring much attention, tiring, and something I would use in the presence of others.
Participants perceived \textit{DwellGestures} ($Med= 4,IQR= 1$) as significantly more accurate than \textit{PursuitsDwell} ($Med= 3,IQR= 1.5$) while seated. When walking, they found \textit{DwellGesture} to be significantly more accurate ($Med= 4,IQR= 1$) and enjoyable ($Med= 4,IQR= 0$) compared to \textit{DwellPursuits} ($Med= 2.5,IQR= 2$, $Med= 3,IQR= 2$), \textit{PursuitsDwell} ($Med= 3,IQR= 1$, $Med= 3,IQR= 2$), and \textit{PursuitsPursuits} ($Med= 2.5,IQR= 1$, $Med= 3,IQR= 2$). \textit{DwellDwell} ($Med= 4,IQR= 2.25$), \textit{DwellGestures} ($Med= 4,IQR= 0$), and \textit{PursuitsGestures} ($Med= 4,IQR= 1$) were perceived as significantly easier to use than \textit{PursuitsPursuits}. 
Table \ref{tab:likert_scale} presents the medians, Interquartile Ranges (IQR) and significance.

\begin{figure} [!t]
    \centering
    \includegraphics[width=\linewidth]{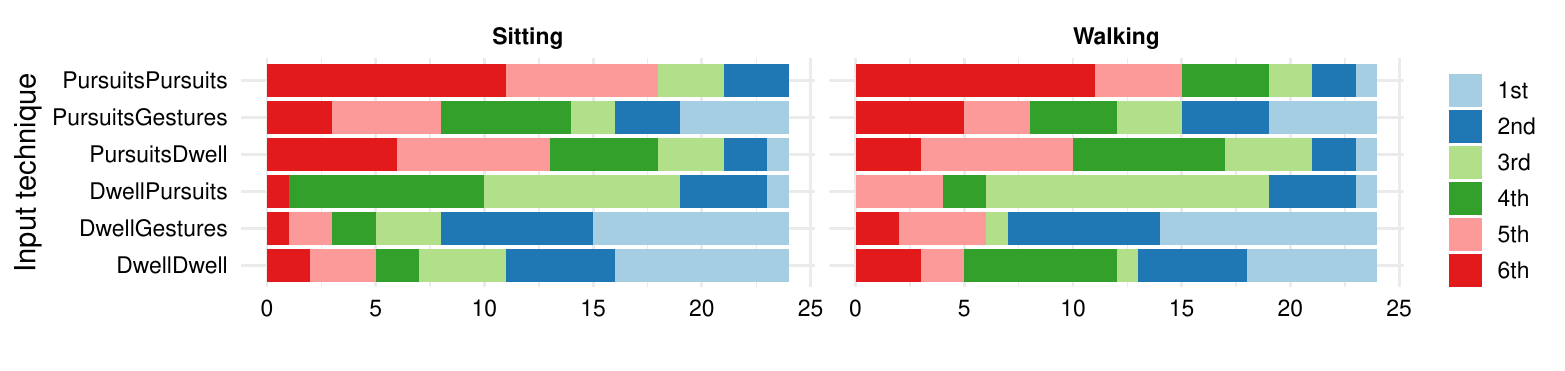}
    \caption{Participants ranked their preference for the input techniques (1= Most preferred; 6= Least preferred). In both seated and walking motor activities, \textit{DwellDwell}, \textit{DwellPursuits}, and \textit{DwellGestures} were consistently preferred, with a significant preference over \textit{PursuitsDwell} and \textit{PursuitsPursuits} while seated. While walking, \textit{PursuitsPursuits} was significantly less favoured than the three.}
    \label{fig:ranking}
\end{figure}

We also asked the participants to express their preference for the proposed input techniques based on their experience  (1= most preferred; 6= least preferred).
In seated motor activities, nine participants identified \textit{DwellGestures} as their most favoured technique, whereas ten participants ranked \textit{DwellGestures} as their top preference in walking scenarios (see Figure \ref{fig:ranking}). Friedman test revealed a significant effect of input techniques on rankings in both activities: sitting, $\chi\textsuperscript{2}(5)$ = 31.5, $p<.0001$ and walking, $\chi\textsuperscript{2}(5)$ = 21.52, $p<.001$, motor activities. Pairwise comparison using Wilcoxon signed-rank tests, adjusted with a Bonferroni correction, showed that \textit{DwellDwell} ($Med= 2, IQR= 3, p<.01$), \textit{DwellGestures} ($Med= 2, IQR= 2, p<.001$), and \textit{DwellPursuits} ($Med= 3, IQR= 1, p<.05$) were ranked significantly better compared to \textit{PursuitsDwell} ($Med= 5, IQR= 1.50$) and \textit{PursuitsPursuits} ($Med= 5, IQR= 1.50$) in sitting condition. 
While walking, \textit{PursuitsPursuits} ($Med= 5, IQR= 2$) was significantly less preferred compared to \textit{DwellDwell} ($Med= 3.50, IQR= 2.25, p<.05$), \textit{DwellGestures} ($Med= 2, IQR= 2.50, p<.005$), and \textit{DwellPursuits} ($Med= 3, IQR= 0.25, p<.01$). \textit{DwellGestures} was also significantly ranked better compared to \textit{PursuitsDwell} ($Med= 4, IQR= 2, p<.05$).

\begin{table}[]
\centering
\caption{While participants perceived \textit{DwellGestures} as significantly more accurate compared to \textit{PursuitsDwell} during seated motor activities, Significant differences were found between multiple pairs on various usability aspects while walking. The numbers represent medians and Interquartile Ranges (IQR), in brackets, along with statistical significance between pairs.}
\label{tab:likert_scale}
\resizebox{\columnwidth}{!}{%
\begin{tabular}{lllllllll}
\hline
                              & \textit{\textbf{DwellDwell (1)}} & \textit{\textbf{DwellGestures (2)}} & \textit{\textbf{DwellPursuits (3)}} & \textit{\textbf{PursuitsDwell (4)}} & \textit{\textbf{PursuitsGestures (5)}} & \textit{\textbf{PursuitsPursuits (6)}} & \textit{\textbf{Friedman test}}                & \textit{\textbf{p \textless .05}}                             \\ \hline
\multicolumn{9}{l}{\textit{\textbf{Sitting}}}                                                                                                                                                                                                                                                                                                                                         \\ \hline
\textbf{Ease of use}          & 4.00 (1.50)                      & 4.00 (1.00)                         & 4.00 (.25)                          & 4.00 (1.00)                         & 4.00 (.50)                             & 4.00 (1.25)                            & $\chi\textsuperscript{2}(5) = 10.22, p> .05$   &                                                               \\ \hline
\textbf{Accuracy}             & 3.00 (2.00)                      & 4.00 (1.00)                         & 4.00 (1.00)                         & 3.00 (1.25)                         & 4.00 (1.00)                            & 4.00 (1.25)                            & $\chi\textsuperscript{2}(5) = 16.85, p< .01$   & (2,4)                                                         \\ \hline
\textbf{Learnability}         & 5.00 (1.00)                      & 5.00 (1.00)                         & 4.00 (1.00)                         & 4.00 (1.00)                         & 4.00 (1.00)                            & 5.00 (1.00)                            & $\chi\textsuperscript{2}(5) = 2.18, p> .05$    &                                                               \\ \hline
\textbf{Enjoyability}         & 4.00 (1.00)                      & 4.50 (1.00)                         & 4.00 (2.00)                         & 4.00 (1.25)                         & 4.00 (2.00)                            & 5.00 (2.00)                            & $\chi\textsuperscript{2}(5) = 14.21, p< .05$   &                                                               \\ \hline
\textbf{Effortlessness}       & 2.00 (2.00)                      & 3.00 (2.00)                         & 2.50 (2.00)                         & 2.00 (1.00)                         & 2.00 (1.25)                            & 3.00 (1.00)                            & $\chi\textsuperscript{2}(5) = 3.85, p> .05$    &                                                               \\ \hline
\textbf{Tiredness}            & 2.00 (2.25)                      & 3.00 (2.25)                         & 3.00 (2.00)                         & 3.00 (1.25)                         & 3.00 (2.00)                            & 2.00 (2.00)                            & $\chi\textsuperscript{2}(5) = 3.37, p> .05$    &                                                               \\ \hline
\textbf{Social Acceptability} & 4.00 (2.00)                      & 4.00 (1.00)                         & 4.00 (1.00)                         & 4.00 (2.00)                         & 4.00 (2.00)                            & 4.00 (1.25)                            & $\chi\textsuperscript{2}(5) = 2.22, p> .05$    &                                                               \\ \hline
\multicolumn{9}{l}{\textit{\textbf{Walking}}}                                                                                                                                                                                                                                                                                                                                         \\ \hline
\textbf{Ease of use}          & 4.00 (2.25)                      & 4.00 (0.00)                         & 3.00 (2.00)                         & 3.00 (2.00)                         & 4.00 (1.00)                            & 2.00 (1.25)                            & $\chi\textsuperscript{2}(5) = 29.15, p< .0005$ & \begin{tabular}[c]{@{}l@{}}(1,6), (2,6),\\ 5,6)\end{tabular}  \\ \hline
\textbf{Accuracy}                      & 4.00 (2.00)                      & 4.00 (1.00)                         & 2.5 (2.0)                           & 3.00 (1.00)                         & 3.00 (1.00)                            & 2.5 (1.00)                             & $\chi\textsuperscript{2}(5) = 23.40, p< .001$  & \begin{tabular}[c]{@{}l@{}}(2,3), (2,4),\\ (2,6)\end{tabular} \\ \hline
\textbf{Learnability}                  & 5.00 (1.00)                      & 5.00 (1.00)                         & 4.00 (.25)                          & 4.00 (1.00)                         & 4.00 (1.00)                            & 4.00 (1.25)                            & $\chi\textsuperscript{2}(5) = 22.63, p< .001$  & (1,6)                                                         \\ \hline
\textbf{Enjoyability}                  & 4.0 (1.00)                       & 4.00 (0.00)                         & 3.00 (2.00)                         & 3.00 (2.00)                         & 4.00 (1.00)                            & 3.00 (2.00)                            & $\chi\textsuperscript{2}(5) = 25.62, p< .001$  & \begin{tabular}[c]{@{}l@{}}(2,3), (2,4),\\ (2,6)\end{tabular} \\ \hline
\textbf{Effortlessness}       & 2.50 (1.00)                      & 2.00 (1.25)                         & 2.00 (.25)                          & 2.00 (1.00)                         & 2.00 (2.00)                            & 2.00 (1.00)                            & $\chi\textsuperscript{2}(5) = 13.67, p< .05$   & (1,6)                                                         \\ \hline
\textbf{Tiredness}                     & 3.00 (2.00)                      & 2.50 (1.25)                         & 4.00 (1.00)                         & 3.50 (1.25)                         & 3.00 (2.00)                            & 4.00 (1.00)                            & $\chi\textsuperscript{2}(5) = 16.26, p< .01$   &                                                               \\ \hline
\textbf{Social Acceptability}          & 4.00 (2.25)                      & 4.50 (2.00)                         & 3.50 (2.25)                         & 4.00 (1.50)                         & 4.00 (2.00)                            & 3.50 (2.25)                            & $\chi\textsuperscript{2}(5) =13.35, p< .05$    &                                                               \\ \hline
\end{tabular}%
}
\end{table}

\subsection{Qualitative Feedback}
We collected feedback through semi-structured interviews that were recorded and transcribed. We employed thematic analysis with an inductive approach to allow us to determine frequently discussed topics by participants. Through this process, the codes were aggregated into subcategories, resulting in three primary themes.

\subsubsection{Theme: Participants' Preference Evaluation:} 
\textit{DwellGestures} was associated with seven positive attributes (n=8), such as intuitive (2), easy to understand (1), natural (1) and fast when Dwell time is paired with gestures, as reported by P16. 
\textit{DwellDwell} was associated with five positive attributes (n=8), such as easy to perform (4) and fast (1). 
Some participants preferred \textit{DwellGestures} and \textit{PursuitsGestures}, attributing this to the gesture part for navigation, as P4 stated that any pair of techniques that include gestures was ranked at the top.
\textit{PursuitsGestures} was associated with five positive attributes (n=5), such as being accurate (1), intuitive (1),  and less tiring to the eyes (1).
P6 reported experiencing eye fatigue when relying on one eye movement (Pursuits). However, they noted that the pairing of Pursuits and gestures appeared to mitigate this issue, potentially reducing eye strain.
Using gestures for navigation was associated with 18 positive attributes, such as being easy for navigation (3), intuitive (3), and accurate (2). It was also explicitly reported as the favoured technique when paired with either Dwell time or Pursuits (5), although it took longer to learn (4). 
P18 appreciated utilising gestures for navigation, deeming it a notably natural action to swipe with their eyes. 
Dwell time, when paired with either Pursuits or Gestures, was associated with five positive attributes (n=5), such as being easier (2) and less effort (1) compared to Pursuits, and natural (1). 
However, Dwell time was less preferred where there were many targets (1).
P4 expressed their preference for Pursuits when paired with other techniques, reporting that any pair of techniques that included Pursuits was ranked higher compared to Dwell time. P18 reported that Pursuits, when used, show a clear interaction. However, Pursuits was also perceived as less accurate (3), tiring to the eyes (2), uncomfortable to look at (1), and not suitable for navigation (1).

\subsubsection{Theme: Participants' Experiences While Mobile:}
Two participants expressed the relevance and comfort of using gaze input while walking.
P12 noted that although using Dwell time for selection was easy when walking, integrating it with gestures for navigation further enhanced the interaction by allowing them to take a break from shifting their gaze between various targets on the screen, as gestures could be performed without the need to focus on a specific target.
While P21 felt that gestures were hard to perform while walking, other participants perceived it more enjoyable (1) and easy (1).
Although \textit{PursuitsGestures} was the favoured technique for P13 during walking, due to the perception of increased accuracy with Pursuits as compared to Dwell time, Pursuits was also viewed as confusing (1), hard (1), and requiring concentration (1). 
In contrast, Dwell time was perceived as better than Pursuits for use while walking (1).
While P19 ranked \textit{DwellDwell} as the most preferred technique for sitting, they ranked \textit{DwellGestures} highest when walking, finding it easier to perform the task with this pair, as the head moves naturally while walking.

\subsubsection{Theme: Contexts and Applications:}
Five participants mentioned using gaze when the temperature is low and wearing impeding clothing, such as gloves; as P10 noted, \say{\emph{I still have to use my nose sometimes, or take my hand out of the gloves, to interact with the phone, which is not perfect}}. They also suggested gaze use when both hands are busy (7), dirty or wet (2), carrying bags (1), or when touch is impossible, such as while driving (1) or cycling (1). Four participants expressed a utilitarian view on gaze input, one stating, \say{\emph{I would have no problems using that anywhere; if it is useful, I would use it}}. Participants reported home use such as reading in bed and swiping pages with gaze (1), listening to music on the couch or bed (1), interacting with hands under a blanket (1), and in stationary situations on public transport (1) or when the device was out of reach (1). Four participants also highlighted accessibility benefits. They also suggested applications to use with gaze, including music players (7), taking photos (2), predefined text selection for typing (1), page navigation (2), album browsing (2), initiating calls (2), and emergency calls (1). While participants were enthusiastic about applications, seven raised concerns about using gaze while walking outdoors.
\section{Discussion \& Future work}

\begin{figure}[t]
    \centering
    \includegraphics[width=\linewidth]{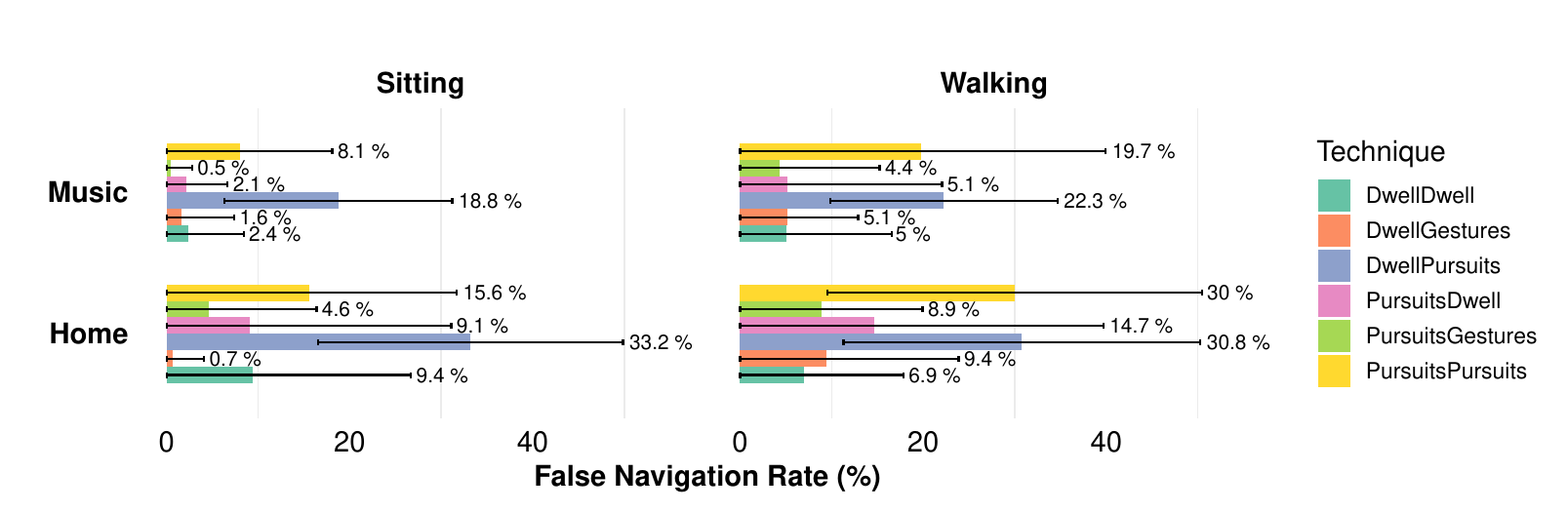}
    \caption{Mean false navigation rate for each navigation technique when paired with other techniques. It is calculated as the proportion of incorrect navigation relative to the total number of navigations performed. Gestures, when paired with Dwell time or Pursuits, resulted in fewer errors compared to Dwell time or Pursuits when used for navigation in sitting motor activity. Pursuits, in \textit{DwellPursuits}, caused the highest navigation error rate. The error bars represent the standard deviation..}
    \label{fig:error_navigation}
    \vspace{-3mm}
\end{figure}

\subsection{Gestures can Enhance Navigation when Paired with Other Techniques}
Our experiment revealed that pairing Pursuits with gestures (\textit{PursuitsGestures}) led to significantly faster task completion times, reduced error rate, and improved completion rates compared to using Pursuits (\textit{PursuitsPursuits}) solely during walking activities. 
The collected feedback showed that participants generally appreciated pairing input techniques rather than using a single eye movement, as one participant explicitly reported that combining gestures with Pursuits reduced eye fatigue caused by using Pursuits only, and another ranked any input technique paired with Pursuits as most preferred. 
Gestures, in particular, when paired with other techniques, was associated with 18 positive attributes.
Likert scale responses also showed that participants perceived gestures as significantly easier to use when paired with Dwell time or Pursuits whilst walking than when using Pursuits only.
Unlike other pairs, we also found no evidence that the motor activity, whether sitting or walking, has an impact on the completion time when using \textit{DwellGestures}.
This pair led participants to rank \textit{DwellGesturees} as the most preferred in both motor activities.
\revision{While pairing gestures with Dwell time and Pursuits positively impacted both techniques, the preference for \textit{DwellGestures} over \textit{PursuitsGestures} could be due to the minimal deliberate eye movements Dwell time requires~\citep{10.1145/2414536.2414609}, its convenience~\citep{10.1145/123078.128728}, and being the gaze counterpart to touch tap~\citep{10.1145/3419249.3420122}.}
Our findings align with prior work that suggested the effectiveness of single stroke gaze gestures~\citep{Møllenbach_Hansen_Lillholm_2013}, the recommendation of gestures with few targets~\citep{10.1145/3544548.3580871}, and its proven robustness while walking~\citep{lei2023dynamicread}.
While complex gaze gestures can cause physiological and cognitive difficulties~\citep{Møllenbach_Hansen_Lillholm_2013,10.1145/1344471.1344477} and was least favoured for selection with many targets~\citep{10.1145/3544548.3580871}, our results show they can be preferred when used to complement rather than replace other gaze input techniques.
In our study, gestures were employed for navigation, a process that P18 perceived as a natural eye-swiping action. This extends prior work that suggested combining multiple techniques, such as Dwell time and Gestures, for issuing different commands~\citep{Møllenbach_Hansen_Lillholm_2013} and the suggested guidelines for enabling users to utilise different techniques based on context and the number of targets~\citep{10.1145/3544548.3580871}.

Based on the results, we emphasise the value of integrating various gaze input techniques in a single user interface. This opens avenues for future research to explore the impact of task complexity and investigate the effectiveness of integrating gaze input techniques for more complex tasks, such as zooming and scrolling. Future work could explore which specific functionalities or commonly used basic gestures~\citep{appleGesturss}, such as zooming, scrolling, or double-tapping, are most effectively executed using particular gaze input techniques. These techniques can be integrated into a single interface, enabling users to familiarise themselves with the functionality of each technique and facilitating the design of an effective mobile gaze-enabled interface that supports fundamental gestures. 
Gaze interfaces may also adapt to individual users, with personalised gaze input techniques paired based on users' performance and preferences.

\begin{figure}[t]
    \centering
    \includegraphics[width=\linewidth]{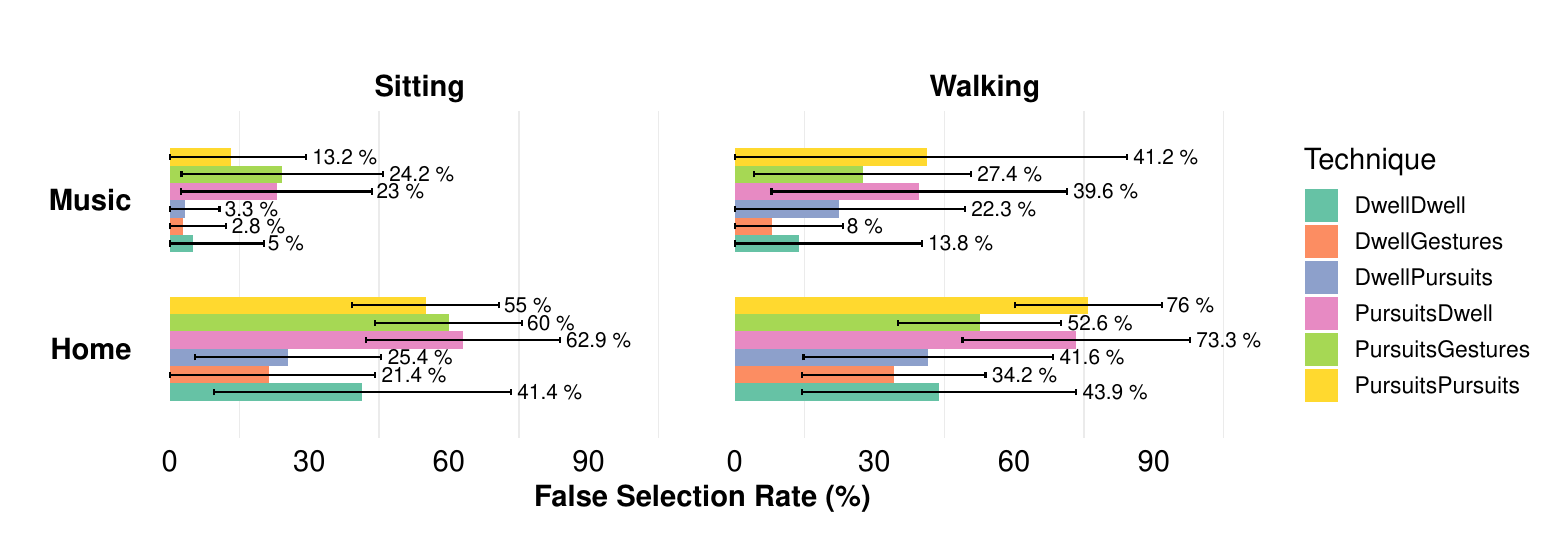}
    \caption{Mean false selection rate for each technique used for selection when paired with other techniques. It is calculated as the proportion of incorrect selections relative to the total number of selections performed. Dwell time, when used for selection and paired with Gestures for navigation (\textit{DwellGestures}, resulted in fewer errors compared to when the technique is used for selection and navigation (\textit{DwellDwell}). The error bars represent the standard deviation.}
    \label{fig:error_selection}
    \vspace{-3mm}
\end{figure}

\subsection{Underlying Mechanisms Influencing Performance}
\begin{figure}[!t]
    \centering
    \includegraphics[width=\linewidth]{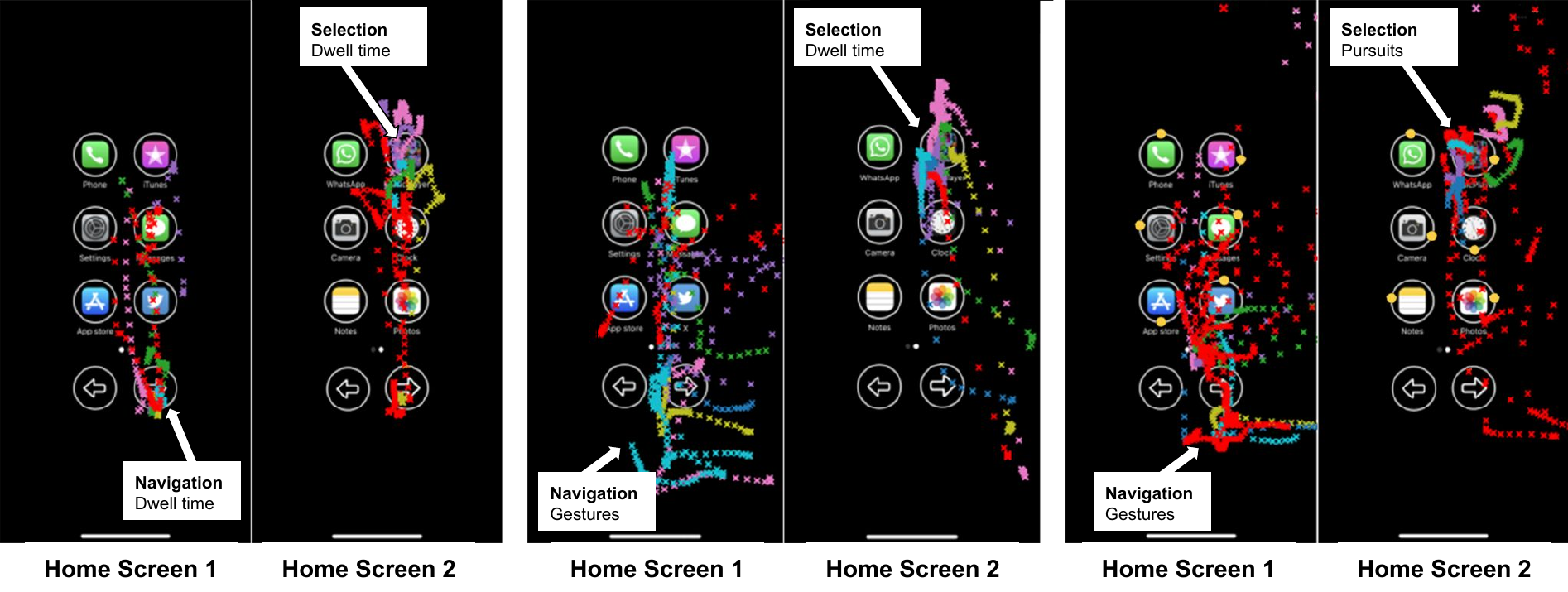}
    \caption{Gaze points from six participants are plotted on the phone's home screen. The red colour represents the gaze points that have led to incorrect actions. Other colours represent the gaze points that ended with each participant's correct actions. \textbf{Left:} participants navigated the home screen using Dwell time and then selected the music player on Home Screen 2 using the same technique. \textbf{Middle:} participants navigated with Gestures and selected the music player using Dwell time. \textbf{Right:} participants navigated with Gestures and selected the music player using Pursuits. 
    }
    \label{fig:screenshort_nav_select}
    \vspace{-3mm}
\end{figure}

Given that Pursuits and gestures both rely on relative eye movements~\citep{khamis2018past, VidalPusuits2013}, and when paired (\textit{PursuitsGestures}), led to a noticeable decline in gesture performance, particularly in terms of task completion time, on both the home and music player interfaces. Regardless of the motor activities, we noticed that Gestures were harder to initiate and took longer to perform (Home screen: $M= 4.0s, SD= 4.07$; Music Player: $M= 13.1s, SD= 6.69$), compared to when Gestures was paired with Dwell time (Home screen: $M= 3.3s, SD= 2.90$; Music Player: $M= 12.3s, SD= 4.65$) for selection. This is likely due to the visual clutter that Pursuits adds to the interface caused by the moving trajectories, which can be visualised in the rightmost screenshots in Figure \ref{fig:screenshort_nav_select}, where red colour crosses in Home Screen 1 show how Pursuits targets were activated while participants were trying to perform the gesture to navigate from Home Screen 1 to 2. However, despite these challenges, Gestures have been shown to be more robust than Pursuits during walking~\citep{lei2023dynamicread}. 
Our experiment supports this by demonstrating that pairing Pursuits with Gestures significantly reduced error rates during walking tasks (Figure \ref{fig:error_rate}). This suggests that although the pairing of Pursuits and Gestures may increase task time, it enhances accuracy and reliability during dynamic activities, such as walking.

When Pursuits is used for navigation and paired with Dwell time for selection (\textit{DwellPursuits}), we noticed an improvement in the navigation speed ($M= 4.59s, SD= 3.66$), but not the error rate in the home screen compared to using Pursuits for navigation in \textit{PursuitsPursuits} ($M= 5.34s, SD= 4.51$). This could be attributed to the inherent speed of the technique~\citep{10.1145/3544548.3580871} while mitigating the negative effects associated with a large number of targets to select from.
To reduce errors and avoid false correlation detection with Pursuits during navigation, which showed the highest error rate compared to Dwell time and Gestures on both home and music player screens in both motor activities (Figures \ref{fig:error_navigation} and \ref{fig:error_selection}), correlation could be computed only when the gaze is within proximity to the Pursuit targets.

Figure \ref{fig:screenshort_nav_select} shows that participants’ gaze often stayed within the same region between screens. For instance, when performing right or left gestures on the home screen (Middle and Right screenshots in Figure \ref{fig:screenshort_nav_select}), the eyes moved to the screen edge and remained there on the next screen. This behaviour suggests UI elements, such as navigation buttons, could remain in consistent areas to support continuity and faster navigation. Placing content where users’ gaze naturally rests can also help maintain focus and reduce search effort.

\subsubsection{Gestures for Navigation Reduces Dwell Unintentional Activation} 
Pairing Dwell time with Gestures for navigation reduced the occurrence of unintentional selections, or the Midas touch effect. This is likely due to the gaze behaviour between screens. This improvement is evident in Figure~\ref{fig:error_selection} and the middle screenshot of Figure \ref{fig:screenshort_nav_select}, where fewer red crosses appear on Home Screen 2, indicating fewer unintentional activations compared to when Dwell time was used alone. In contrast, the left-hand screenshots of Figure \ref{fig:screenshort_nav_select} show a higher number of accidental dwell target activations when participants shifted their gaze from the navigation button to the music player. The mean false selection rate in the \textit{DwellDwell} condition was $41.39\%$ ($SD= 31.9$) on the Home screen, and $5.03\%$ ($SD= 15.3$) in the music player, while this error reduced to $21.37\%$ ($SD= 22.7$) on the home screen and to $2.78\%$ ($M= 9.4$) in the music player when using Dwell time for selection in the \textit{DwellGesture} (see Figure \ref{fig:error_selection}).

\subsection{Number of Targets, Size, and Placement Affect Gaze-Based Technique Performance}
While pairing different techniques proved promising, some pairings were not as promising. 
While this could be due to the nature of the task, we suspect three other factors may be involved, which we discuss in the following sections.

\subsubsection{The Effect of Number of Targets}
The effect of the number of targets may impact the accuracy and completion time of paired techniques in various degrees, as suggested by prior work~\citep{10.1145/3544548.3580871}. For example, when walking, the experiment data showed that \textit{PursuitsGestures} is less accurate when used in the home screen, with an average error rate $50.26\%$ (See Figure \ref{fig:screenshot}), than when used in the music player application ($M= 15.35\%$). \textit{PursuitsDwell} also resulted in more errors compared to \textit{DwellPursuits}, as the former displayed more targets to be selected using Pursuits, while the latter pair used Pursuits for navigation with only two targets (see Figures \ref{fig:error_navigation} and \ref{fig:error_selection}). One approach to mitigate the impact of the number of targets is to display a limited number of functions or targets on each screen. This would allow users to concentrate on a more concise set of tasks, potentially enhancing their overall performance.

\subsubsection{The Effect of Target Size}
In our implementation, we ensured that the target size remained consistent across all the input techniques employed, following Apple design guidelines and inspired by prior work on gaze interaction on mobile devices~\citep{10.1145/3544548.3580871, apple_guide}. While this approach guarantees a balanced comparison among the techniques, it impacts the performance of each technique to varying extents. This issue could be mitigated by designing the interface with specific consideration for the input techniques, where targets with Dwell time, for example, can be sized to better compensate for tracking inaccuracy~\citep{10.1145/3025453.3025599, 10.1145/3706598.3713092}. 

\subsubsection{The Effect of Target Placement}
The placement of the targets on the screen may also impact the performance of the input techniques, as this was suggested by prior work~\citep{10.1145/3544548.3580871, 10.1145/3025453.3025599, 10.1145/3706598.3713092}. 
In our implementation, we endeavoured to position all targets near the centre of the screen to minimise such an effect. However, some targets were placed closer to the edges due to limitations on screen real estate. Although randomising the target positions might mitigate such effects in the study, it could introduce a time cost due to prolonged search times, potentially affecting the analysis. In designing a gaze-enabled user interface, careful attention must be given to the placement of targets, ensuring they align with the variations in accuracy and precision across different screen regions~\citep{10.1145/3544548.3580871, 10.1145/3025453.3025599}.

\subsection{Performance in Real-world Mobile Usage}
\revision{By conducting our experiment in a corridor instead of a lab-controlled setting, we introduced uncontrollable variables such as the passersby.
While this approach may contribute towards understanding the real-world applicability of our findings, outdoor or in-the-wild studies provide more realistic conditions due to the inherent environmental and contextual factors such settings introduce ~\citep{10.1145/1409240.1409253, Chittaro2010}. Issues include changes in the weather, the various lighting conditions, the shaky environments due to the dynamic use, disturbance by pedestrians or obstacles, the limited or split attention, and the added distractibility by public events~\citep{10.1145/1409240.1409253, 10.1145/3229434.3229452, Chittaro2010}.
As mentioned in Section \ref{sec:limitation}, we intentionally chose an indoor environment, with minimal distraction, and controlled lighting conditions to explore the performance of pairing gaze-based techniques in isolation. Our findings provide an initial step toward future investigations of pairing gaze techniques in the wild.}

\revision{To address issues in a shaky environment, which can often break calibration data, eye-tracking systems could leverage IMU sensor data to detect when recalibration is needed, depending on the user's context, as done in MAC-GAZE~\citep{lei2025macgazemotionawarecontinualcalibration}. 
As explicit calibration is tedious and time-consuming~\citep{10.1145/3229434.3229452}, implicitly calibrating users is promising~\citep{10.1145/3706598.3713739, 9802919}, especially for the mobile context, where users are likely to calibrate more often due to the dynamic nature of mobile use~\citep{10.1145/3229434.3229452}.}

\revision{While Eyedid SDK, the eye-tracking library used in our experiment, utilises RGB images for gaze estimation, RGB cameras are sensitive to lighting conditions~\citep{10.1145/3229434.3229452, majaranta2014eye}. For example, tracking eyes using RGB cameras in the dark is challenging.
Similar to Eyedid, ASGaze proposed using RGB cameras to enable eye tracking based on geometric eye models~\citep{10.1145/3560905.3568544}.
While avoiding the high cost of commercial solutions, ASGaze enables tracking on various external surfaces, such as non-electronic whiteboards and computer screens, and is not limited to phone screens. 
When tested under various settings, ASGaze achieved good tracking performance, but its performance degrades when illumination is very low, as it becomes harder to extract the iris boundaries reliably. The tracking error with ASGaze also increases with the user's walking speed. This could be due to increased motion blur and larger variations between consecutive frames caused by hand-induced phone shaking, affecting iris boundary extraction, a key component of eye tracking in ASGaze.
On the other hand, the use of the Infrared-based Pupil-Corneal Reflection (IR-PCR) mitigates tracking issues in the dark but comes at the cost of not being reliable under direct sunlight, as sunlight may interfere with the IR sensors~\citep{10.1145/3229434.3229452, majaranta2014eye}.}

\revision{Moving from frame-based cameras to event-based cameras for tracking may be a plausible approach to take, given that conventional frame-based cameras are limited by the fixed rate sampling, often bounded by a few hundred hertz, depending on the tracking hardware~\citep{NEURIPS2023_c41b5d8c}.
Recent system EV-Eye~\citep{NEURIPS2023_c41b5d8c} utilised a bio-inspired event camera to capture asynchronous, pixel-level intensity changes with sub-microsecond latency, enabling adaptive and high-frequency eye tracking. In EV-Eye, each pixel asynchronously reports brightness changes with precise timestamps, enabling continuous, low-latency representation of motion without using frames.
As event cameras produce events only when intensity changes are detected, they save sensing energy and bandwidth~\citep{NEURIPS2023_c41b5d8c}. Given their high dynamic range (140 dB compared to 60 dB for traditional RGB cameras), such cameras operate effectively under challenging lighting conditions, making them suitable for eye tracking on handheld mobile devices, where lighting conditions are likely to vary.}

\subsection{Gaze Has Potential for Addressing Situational Impairments for Mobile Users} \label{sec:situational_impairment}
Participants suggested using gaze for interaction in contexts or situations that may negatively affect their experience when interacting with mobile phones. While walking is one of these factors, they suggested using gaze when the device is out of reach, when hands are encumbered with various objects such as shopping bags~\citep{10.1145/3152771.3156161}, when hands are busy, or when wearing gloves and the weather is cold. These factors were previously identified by Wobbrock et al.~\citep{wobbrock2019situationally} as contextual factors that can negatively affect user interaction with mobile devices, leading to situationally induced impairments.
Given that situational impairments can impact users of various abilities~\citep{wobbrock2006future, 10.1145/3152771.3156161,wobbrock2019situationally}, as users experiencing situational impairments might interact with their touch devices in ways similar to individuals experiencing hand tremors,
it is crucial to explore design solutions to accommodate such impairments and mitigate their impact. 
Although we did not optimise the interface to minimise the effect of walking in our experiment, as it was out of scope, our results offer insight into how each technique pairing performs and is perceived by participants while walking. Future work could investigate different ways the interface can be designed, including what gaze input technique to use for specific situations and functionalities to account for the complexity present during mobile interaction. Solutions can be borrowed from prior work that evaluated the use of touch for interaction when walking~\citep{10.1145/1409240.1409253} and the presented preliminary design guidelines to account for situational impairments~\citep{saulynas2022putting}.

\section{Conclusion}
In this work, we conducted a study to investigate the impact of pairing gaze input techniques, namely Dwell time, Pursuits, and gestures, within the same interface on the usability of navigation and selection tasks while interacting with mobile devices during two motor activities: sitting and walking. 
Our results demonstrate that pairing techniques can generally enhance the effectiveness of gaze interaction on mobile devices, particularly for navigation and selection tasks. While gestures generally improved performance when combined with Dwell time or Pursuits, \textit{PursuitsGestures} resulted in significantly faster completion times, reduced error rates, and improved completion rates when walking compared to Pursuits only. Participants reported better performance and ease of use compared to using Pursuits only.
In both motor activities, \textit{DwellGestures} was the most favoured technique by the participants. 
Our results show several future directions to explore the impact of combining gaze techniques for more complex tasks. We also discussed the implications of combining gaze-based interactions in the same interface and investigated why certain combinations work better than others. Our work opens new directions for enhancing gaze-based interactions on mobile devices and better leveraging gaze input for interacting with mobile devices.
\section*{Acknowledgements and Funding}
This work was supported by the Islamic University of Madinah. 
We acknowledge the use of ChatGPT-5 to enhance the readability and shorten certain parts of the manuscript through paraphrasing. It was not used for the entire manuscript.

\section*{Author Contributions}
CRediT: \textbf{Omar Namnakani:} Conceptualization, Data curation, Formal analysis, Funding acquisition, Investigation, Methodology, Project administration, Software, Validation, Visualization, Writing – original draft, Writing – review \& editing; \textbf{Yasmeen Abdrabou:} Conceptualization, Methodology, and Writing – review \& editing; \textbf{Jonathan Grizou:} Supervision; \textbf{Mohamed Khamis:} Conceptualization, Methodology, Resources, Supervision, Writing – review \& editing. 

\section*{Disclosure Statement}
The authors report there are no competing interests to declare.

\bibliographystyle{apacite}
\bibliography{interactapasample}


\end{document}